\documentclass[preprint,showpacs,preprintnumbers,amsmath,amssymb]{revtex4}
\usepackage{graphicx}
\usepackage{dcolumn}
\usepackage{bm}
\usepackage{color}

\begin{document}
\title{Twisting short dsDNA with applied tension}

\author{Marco Zoli}

\affiliation{School of Science and Technology \\  University of Camerino, I-62032 Camerino, Italy \\ marco.zoli@unicam.it}

\date{\today}

\begin{abstract}
The twisting deformation of mechanically stretched DNA molecules is studied by a coarse grained Hamiltonian model incorporating the fundamental interactions that stabilize the double helix and accounting for the radial and angular base pair fluctuations. The latter are all the more important at short length scales in which DNA fragments maintain an intrinsic flexibility. The presented computational method simulates a broad ensemble of possible molecule conformations characterized by a specific average twist and determines the energetically most convenient helical twist by free energy minimization. As this is done for any external load, the method yields the characteristic twist-stretch profile of the molecule and also computes the changes in the macroscopic helix parameters i.e. average diameter and rise distance. It is predicted that short molecules under stretching should first over-twist and then untwist by increasing the external load. Moreover, applying a constant load and simulating a torsional strain which over-twists the helix, it is found that the average helix diameter shrinks while the molecule elongates, in agreement with the experimental trend observed in kilo-base long sequences.  The quantitative relation between percent relative elongation and superhelical density at fixed load is derived. The proposed theoretical model and computational method offer a general approach to characterize specific DNA fragments and predict their macroscopic elastic response as a function of the effective potential parameters of the mesoscopic Hamiltonian.
\end{abstract}


\maketitle

\section*{I. Introduction}

Understanding the DNA mechanics has important implications as, in cells, DNA is constantly bent, stretched, repaired and processed by proteins which, upon binding, confer to the double helix its biological functions and regulate gene expression \cite{stasiak,forget,cappello,wang12,albu14a}. 

The development of optical and magnetic tweezers techniques over the last twenty five years has allowed to gain remarkable insights into the elastic properties of single DNA molecules by studying their response to external forces in the pico-Newton regime \cite{chu,busta92,busta94,cluzel96,block97,mameren09}. Such forces are required to oppose the thermal bending fluctuations due to the environment which constantly buffet the molecular bonds thus causing the helix to assume different random walk configurations. In fact, at physiological temperatures, the thermal energy per nano-meter is $\sim 4 pN$. 

Force-extension data of kilo base pairs sequences have been well reproduced by worm-like-chain models that treat DNA as an isotropic rod whose behavior is dominated by entropic elasticity at least up to $\sim 10 pN$ \cite{busta00}. Instead, at higher external forces, structural changes occur in the intra-strand base pair covalent bonds and the helix is progressively over-stretched to a length larger than its B-form contour length \cite{rouz01}.
Later measurements by the rotor bead tracking technique \cite{busta06} have shown that kilo base pairs DNA molecules over-twist upon stretching up to $\sim 30 pN$ and then untwist above such value.   

Importantly, if a torque is applied in order to over-twist the double helix \textit{under a constant load}, it has also been found that the molecule extends. While analogous results have been obtained by magnetic tweezers experiments, molecular models for DNA in a solvent \cite{croq06,zachar15} have suggested that a negative inclination of the base pairs towards the minor groove could reduce the helix diameter and elongate the rise distance. Moreover, the molecule extension appears to be a linear function of the applied over-twist in the limited range of those torsional strains which preserve the stable B-form.

Motivated by these findings pointing to a remarkable DNA flexibility together with a rich interplay between its twisting and stretching properties \cite{marko97,wuite11},  we have developed in a previous work \cite{io17}  a computational method  based on a mesoscopic DNA Hamiltonian which treats the helix at the base pair level and retains the fundamental intra-strand and inter-strand base pair interactions, responsible for the helix stability in the presence of a solvent.  Mesoscopic models have the capability to predict the thermomechanical behavior of specific sequences through optimization of the potential parameters via direct fitting of experimentally accessible data e.g., the melting profiles \cite{campa,albu14,weber15a,singh15}.  
Essentially our method assumes that the single molecule may exist in a broad range of helical conformations, specified through the average number of base pairs per helix turn, and determines the energetically most convenient conformation by free energy minimization.

As the computation is carried out by varying the strength of an external load, one can predict the twisting response of the molecule as a function of the stretching perturbation. While the method has been applied to short fragments which have been the focus of recent and widespread interest in view of their unexpected flexibility \cite{widom,volo05,olson06,archer,fenn,mastro,gole,vafa,kim14,mazur14,tan15,io16a,zhen17,tan16},  the same scheme can be used (compatibly with the available CPU time) for any sequence and length being aware that the latter generally affect the properties of the molecules \cite{volo10,olson10}.

In this paper, going beyond our previous study, we assume that the helix may be over-twisted (or untwisted) with respect to its equilibrium conformation \textit{under a constant load} and investigate the ensuing modification on the helical shape. 
{It is emphasized that the load has here the function to align, not that to disrupt \cite{piana05}, the intra-strand stacking bonds. Accordingly, the external force is tuned within a range of values which do not cause the over-stretching of the molecule backbone.  In this way, the method offers a feasible approach to simulate the above described experimental setup.}
In particular, we derive here the quantitative relation between average helical elongation and superhelical density which, in principle, could be investigated experimentally for sequences of a few tens of base pairs. Furthermore, it is shown that the over-twisting / untwisting transition, observed in kilo-base long sequences as a function of the external force, is predicted by our model and essentially ascribed to a dependence of the helix bending fluctuation on the size of the applied load.

The geometrical representation for the helix is outlined in Section II while the mesoscopic Hamiltonian model is discussed in Section III. The general features of the computational method are given in Section IV and the formulas for the macroscopic helix parameters are defined in Section V. The results are presented in Section VI while some conclusions are drawn in Section VII.

\section*{II. Helical Model  }

\begin{figure}
\includegraphics[height=8.0cm,width=8.0cm,angle=-90]{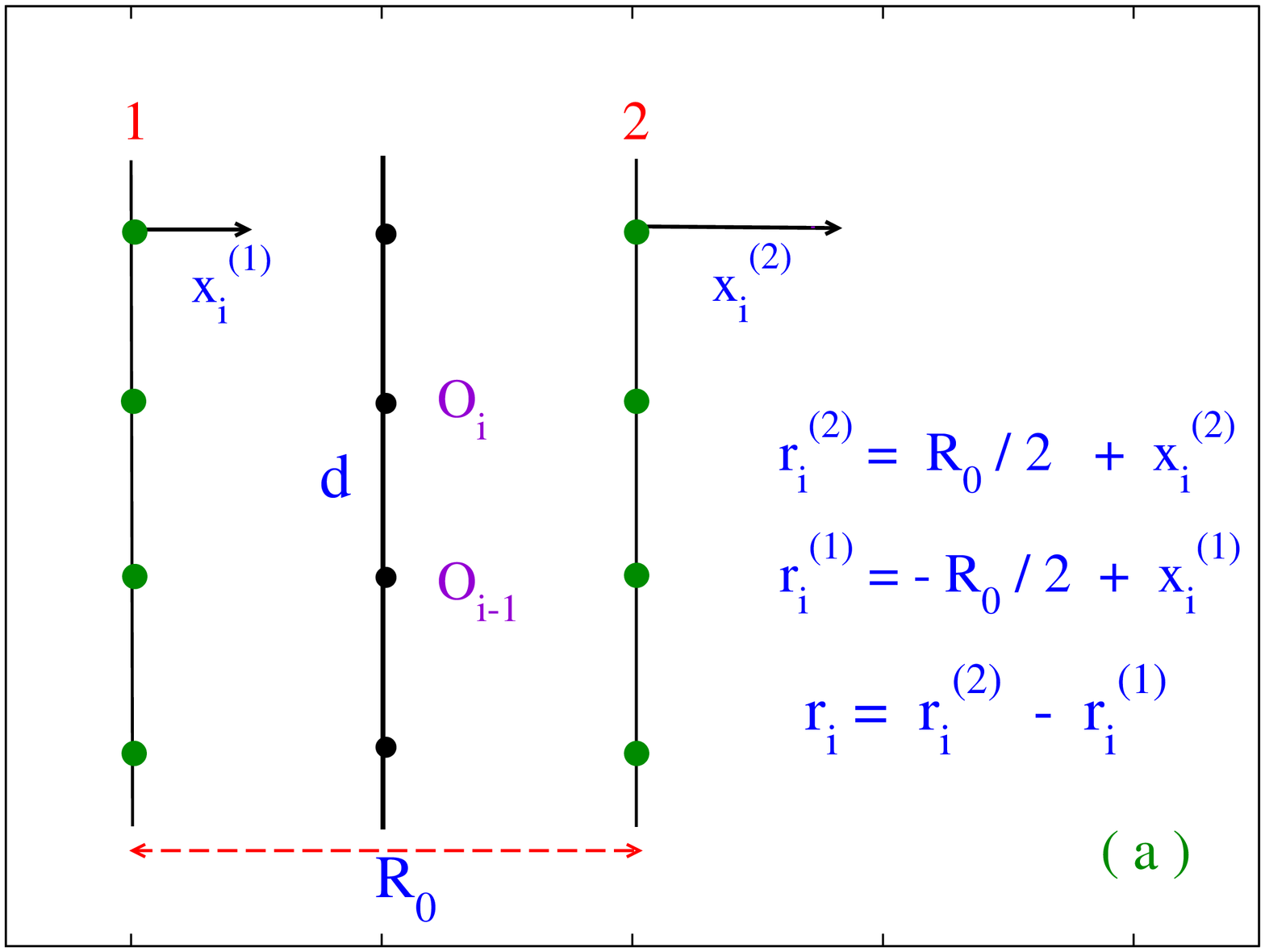}
\includegraphics[height=8.0cm,width=8.0cm,angle=-90]{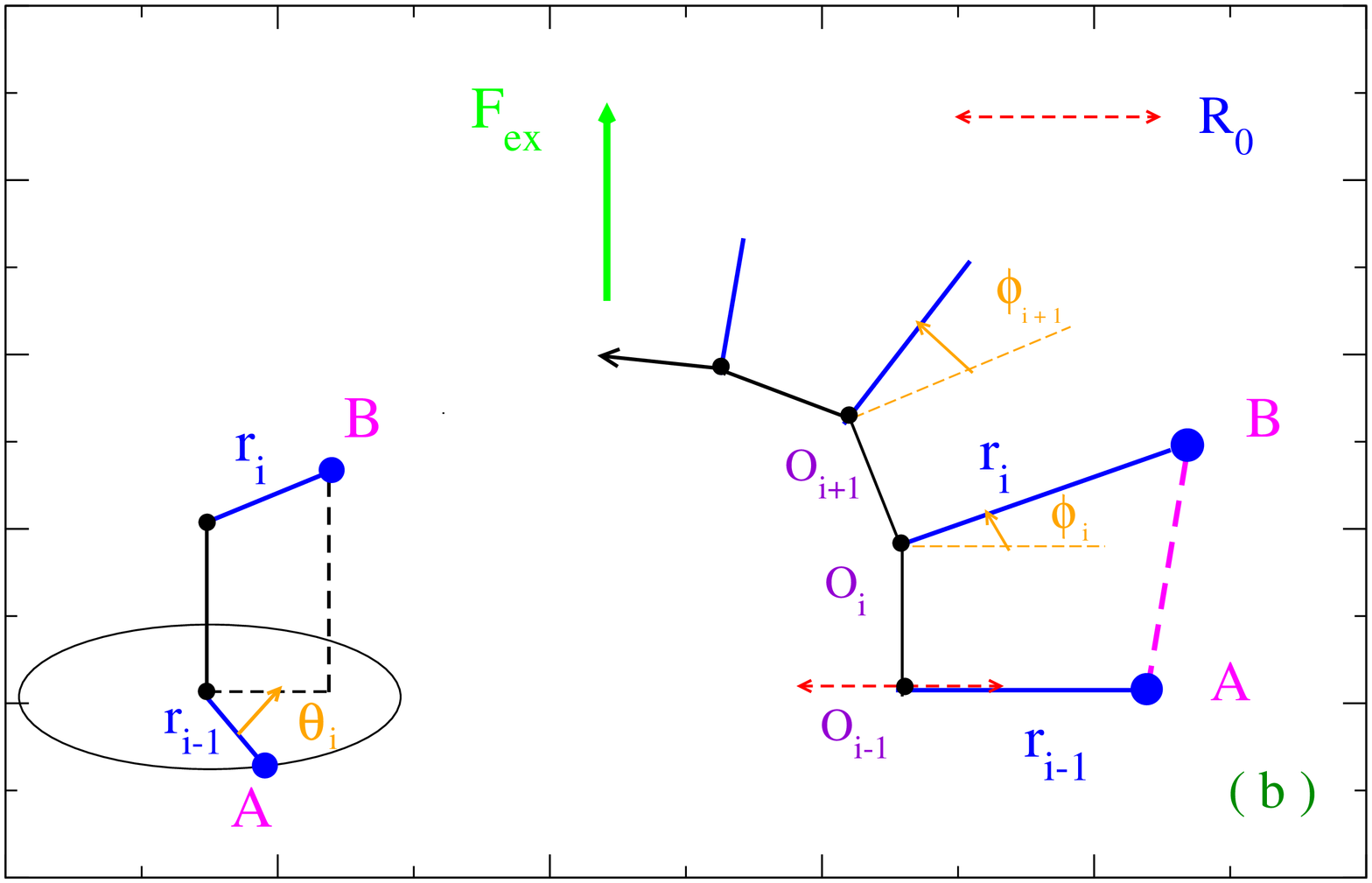}
\caption{\label{fig:1}(Color online)  
(a) Schematic of a simple ladder model for N base pairs whose mates are arranged along the complementary strands. The pair mates vibrate around their respective positions on the strands which are separated by the helix diameter $R_0$. The relative base pair distance $r_i$ is measured with respect to the central molecule axis. 
{$d$ is the rise distance along the molecule backbone in the absence of fluctuations.} (b) In a more realistic picture for the helix, adjacent base pairs along the stack are twisted and bent. Left panel: $\theta_i$ is the torsional angle between $r_i$ and $r_{i-1}$. Right panel: $\phi_i$ is the bending angle between $r_i$ and $r_{i-1}$. The molecule backbone lies on the sheet plane and it is subject to an external load $F_{ex}$ which also lies on the sheet plane,  {it is applied to the molecule backbone } and it is set along the direction of the rise distance $\overline{O_{i}O_{i-1}}$.
}
\end{figure}

In previous Hamiltonian studies of DNA denaturation \cite{pey04}, the double helix has been described by a basic ladder model, see Fig.~\ref{fig:1}(a), in which the bases are arranged as beads along the complementary strands. The backbone of a molecule with $N$ base pairs is thus a chain of $N-1$ segments connecting the points $O_i \,{}\, (i=\,1,...,N) $.  {$R_0$ and $d$, input parameters of the model, represent the bare helix diameter and rise distance along the molecule stack, respectively, in the absence of fluctuations.} 
Each pair is formed via the hydrogen bond connecting the two mates and only two degrees of freedom per pair, $x_{i}^{(1)}$ and $x_{i}^{(2)}$, representing the displacements of the pair mates, are included in the model.  The in-phase-displacement, $x_{i}^{(1)} + x_{i}^{(2)}$, yields a straightforward harmonic potential energy term in the chain Hamiltonian which can be exactly integrated. Instead the relative distance, $r_i=\,x_{i}^{(1)} - x_{i}^{(2)}$, measured from the central helical axis, stretches the hydrogen bond and determines the statistical mechanics of the DNA ladder model \cite{io09,io10}.  In fact, $r_{i}$ may even become smaller than $R_0$ thus compressing the hydrogen bond but the pair mates cannot get too close to each other due to the strands repulsion exerted by the negatively charged phosphate groups \cite{tan08}.

Here we adopt a more realistic picture for the double helix which goes beyond the ladder model assuming that adjacent displacements along the molecule stack, e.g. $r_{i}$ and $r_{i-1}$, are allowed to twist and bend as shown in Fig.~\ref{fig:1}(b). Accordingly, the distance $\overline{AB}$ between neighbor base pairs is a function of the rotational degrees of freedom with both the torsional angle $\theta_i$ and the bending angle $\phi_i$ being integration variables in the calculation of the partition function.
Hence twisting and bending fluctuations are incorporated in our model whereas other structural deformations, such as propeller twist that enhances the intra-strand base pair stacking and the presence of grooves relevant to the sequence specific DNA-protein binding \cite{orozco15,hwang16}, are not taken into account. 
{ While a general description of the base pair degrees of freedom and helical parameters is given e.g., in Ref. \cite{olson01}, the effects of heterogeneous base pair sequence are analyzed e.g. in a comprehensive molecular dynamics simulation for a large set of oligomers \cite{lavery10}. }

{ Restricting our study to homogeneous fragments, we further assume that,} for short DNA molecules, the most energetically convenient conformations are those for which the helix axis is essentially planar \cite{bates,irob}, the $O_i$'s are pinned to the sheet plane.  This is consistent with the fact that, once short molecules close into a ring, the free energy of supercoiling is mostly partitioned into twisting while the writhe contribution to the linking number is negligible  \cite{shore,horowitz}.

{As the base pairs are described by the $r_i$'s which depart from the central molecule axis, we assume: \textit{1)} to apply the force $F_{ex}$ to such axis, along the direction of the segment $\overline{O_{i}O_{i-1}}$ in Fig.~\ref{fig:1}(b), \textit{2)} that the force pulls one end of the chain while the opposite end remains anchored,  \textit{3)} that, for short molecules, the force acts uniformly on all the chain segments and \textit{4)} that the force has the effect to orient the chain segments along a specific direction. More generally, in a two strand representation and in the experimental setup, one can devise a number of different schemes to pull the chain ends applying the load either to one or both complementary strands \cite{seno10,romano13,fiasconaro13}. Moreover, even the pulling direction may become important yielding specific deformations of the helical structure, as shown by molecular force balance experiments on heterogeneous duplexes. This occurs if the forces and loading rates are such to disrupt the intra-strand stacking bonds \cite{gaub08}. These latter cases however do not pertain to our analysis as emphasized in the Introduction. 
We instead assume that $F_{ex}$ varies in a range of low to moderate values which do not over-stretch the helix and study the response of the homogeneous molecule to the external perturbation.    }

As detailed in the next Section, the computational technique sums over a large ensemble of molecule configurations and finds by free energy minimization \textit{a)} the average twist fluctuation i.e., the average number of base pairs per helix turn, \textit{b)} the average radial fluctuation which provides a measure of the molecule diameter and \textit{c)} the average intra-strand elongation. This allows to determine the twisting profiles and the molecule structural deformations as a function of $F_{ex}$.

\section*{III. Hamiltonian Model}

Mesoscopic models and statistical mechanics methods have been widely employed over the last decades to study both thermal equilibrium and dynamical properties of DNA 
\cite{wart,zhang,benham99,joy05,zdrav06,krueg,kalos09,pey09,rapti10,bishop14,weber15,singh16}. A major research focus has been the denaturation transition with the associated Watson-Crick base pair openings and formation of bubble profiles which are crucial to the DNA biological functioning \cite{bonnet03,zocchi04,rapti06,kalos11,hand12,metz12,palm13}. Theoretical analysis of the helix melting transition depart either from Ising-like models describing paired and unpaired complementary bases \cite{poland}  or from Hamiltonian models treating the hydrogen bonds through a potential energy function of the distance between the pair mates \cite{proho85,pey89}. While Ising models for DNA melting have been developed essentially for long chains in which loops with hundreds of open base pairs largely contribute to the partition function, mesoscopic Hamiltonian models have proven effective to deal with short DNA sequences whose dynamics is dominated by strong base pair fluctuations. Thus, for chains of order $\sim 100$ base pairs, it is desirable to describe the intermediate fluctuational states (between the closed and open base pair configurations) in terms of a continuous variable, i.e. the relative separation between the pair mates on complementary strands. Furthermore, stability properties and flexibility of the helix can be studied in Hamiltonian models by introducing stacking potentials which account for the covalent bonds between neighbor nucleotides along the molecule strands. Certainly, inter-strand and intra-strand forces cannot be treated independently as the disruption e.g., of the \textit{i-th} hydrogen bond moves one (or both) mate(s) of the \textit{i-th} base pair out of the stack thus affecting also the intra-strand interactions between the adjacent $i$ and $i \pm 1$ bases.

These requirements have been met by the Hamiltonian model used in recent analysis of the flexibility of short DNA sequences which have predicted cyclization values in fair agreement with available experimental data \cite{io16b}.  The same Hamiltonian, consistent with the helical model of Fig.~\ref{fig:1}(b), is adopted in the present study.
In addition, we introduce the effect of a tunable external load which stretches the molecule axis and induces the changes in the optimal twisting configuration as determined by minimization of the free energy.

Then, the Hamiltonian for the helical molecule with $N$ base pairs of reduced mass $\mu$, stretched by a force $F_{ex}$, is:

\begin{eqnarray}
& &H =\, H_a[r_1] + \sum_{i=2}^{N} H_b[r_i, r_{i-1}, \phi_i, \theta_i] \, , \nonumber
\\
& &H_a[r_1] =\, \frac{\mu}{2} \dot{r}_1^2 + V_{1}[r_1] \, , \nonumber
\\
& &H_b[r_i, r_{i-1}, \phi_i, \theta_i]= \,  \frac{\mu}{2} \dot{r}_i^2 + V_{1}[r_i] + V_{2}[ r_i, r_{i-1}, \phi_i, \theta_i] - F_{ex} d  \cos\bigl( \sum_{k=1}^{i-1}\phi_k \bigr)  \, . \nonumber
\\ 
\label{eq:01}
\end{eqnarray}

Note that $H_a[r_1]$ is treated separately as the first base pair has no preceding neighbor along the molecule backbone.

$V_{1}[r_i]$ is one-particle potential accounting for the hydrogen bond between the \textit{i-th} pair mates and $V_{2}[ r_i, r_{i-1}, \phi_i, \theta_i]$ is the two-particle stacking term which also depends on the angular degrees of freedom.

Their analytic expressions are:

\begin{eqnarray}
& &V_{1}[r_i]=\, V_{M}[r_i] + V_{Sol}[r_i] \, , \nonumber
\\
& &V_{M}[r_i]=\, D_i \bigl[\exp(-b_i (|r_i| - R_0)) - 1 \bigr]^2  \, , \nonumber
\\
& &V_{Sol}[r_i]=\, - D_i f_s \bigl(\tanh((|r_i| - R_0)/ l_s) - 1 \bigr) \, , \nonumber
\\
& &V_{2}[ r_i, r_{i-1}, \phi_i, \theta_i]=\, K_S \cdot \bigl(1 + G_{i, i-1}\bigr) \cdot \overline{d_{i,i-1}}^2  \, , \nonumber
\\
& &G_{i, i-1}= \, \rho_{i, i-1}\exp\bigl[-\alpha_{i, i-1}(|r_i| + |r_{i-1}| - 2R_0)\bigr]  \, . \nonumber
\\ 
\label{eq:02}
\end{eqnarray}

The one-particle potential comprises: \textit{a)} a hydrogen bond Morse potential $( V_{M}[r_i] )$ for the $i-th$ base pair with spatial range $b_i$ and dissociation energy $D_i$. \textit{b)} A solvent potential $( V_{Sol}[r_i] )$ which enhances the threshold for base pair dissociation and stabilizes the hydrogen bond through the parameters $f_s$ and $l_s$ as pointed out in refs. \cite{druk,io11}.   

The two-particle potential depends on the distance $\overline{d_{i,i-1}}$ between adjacent $r_i \,, r_{i-1}$, as marked by the $\overline{AB}$ segment in Fig.~\ref{fig:1}(b), which includes the angular variables.
The stacking potential contains both an elastic force constant $ K_S$ and  nonlinear terms weighed by the parameters $\rho_{i, i-1}$, $\alpha_{i, i-1}$ which favor cooperative behavior in the formation of local bubbles as it is understood by the following observation. In fact,  when the conditions $|r_i| - R_0 \ll  \alpha_{i, i-1}^{-1}$ and $|r_{i-1}| - R_0 \ll  \alpha_{i, i-1}^{-1}$ are fulfilled,  the effective stacking coupling is  { $\sim  K_S \cdot \bigl(1 +  \rho_{i, i-1}  \bigr)$ }
hence, both the $i$ and $i-1$ base pairs are bound. However, thermal fluctuations may violate either-or of the above conditions thus weakening the relative hydrogen bond. In this case, $G_{i, i-1} \rightarrow 0$, the corresponding coupling drops to $\sim  K_S$ and also the adjacent base along the stack loosens its hydrogen bond. Accordingly, the fluctuational opening spreads along the strands and untwists the helical molecule. This qualitative picture indicates that the equilibrium twist conformation is sensitive to the interplay between stacking parameters and amplitude of the base pair separations.

For homogeneous sequences, one can shorten, $\rho_i \equiv \, \rho_{i, i-1}$, $\alpha_i \equiv \, \alpha_{i, i-1}$. Then, consistently with our previous works \cite{io16b} we take in the following calculations,  $D_i=\,40 meV$, $b_i= 5 \AA^{-1}$,  $f_s=\,0.1$, $l_s=\,0.5 \AA$,  $K_S=\,10 mev \AA^{-2}$, $\rho_{i}=\,1$, $\alpha_{i}=\, 2 \AA^{-1}$.  {For the bare helix diameter and rise distance, we set the values $R_0 = \,20 $\AA {} and $d = \, 3.4$\AA {}, respectively.}
This set of model parameters reproduces the experimentally estimated DNA free energies per base pair  \cite{krueg,metz11}. Further details on the potentials in Eq.~(\ref{eq:02}) are given e.g., in refs. \cite{io11,io12}. 

{The fourth addendum in the last of Eq.~(\ref{eq:01}) accounts for the presence of the external force field as described in Section II. 
As $F_{ex}$ has the purpose to straighten the chain thus opposing the coiling effect of the bending fluctuations, we expect that by increasing $F_{ex}$, the entropy is reduced and, accordingly, the free energy grows. }

{ It is also worth mentioning that our mesoscopic Hamiltonian may be extended to study the thermodynamics and flexibility properties of RNA \cite{dekker14,tan17} provided that a specific parametrization can be obtained for a model potential with twisting and bending variables.}

The Hamiltonian in Eq.~(\ref{eq:01}) basically represents a DNA molecule, in a solvent potential and stretched by an external load, whose equilibrium statistics is obtained by solving the associated partition function. This task is carried out by generating an ensemble of molecule configurations consistent with the model potential and integrating, for each base pair in the sequence, over a distribution of radial and angular degrees of freedom dense enough to stabilize the partition function. Previous studies based on molecular dynamics, Monte Carlo and transfer integral techniques for the Hamiltonian of the simplified ladder model in Fig.~\ref{fig:1}(a) \cite{pey2,ares,zhang}  have encountered the problem to define an upper bound for the integral over the base pair distances $|r_i|$. The problem stems from the fact that the two-particle potential of the ladder model vanishes for the zero mode (all $r_i$'s equal) whereas the Hamiltonian remains finite for $r_i \rightarrow \infty $ as the one-particle potential is always bounded. Since $V_{1}[r_i]$ is not translationally invariant, the zero mode cannot be removed and the partition function diverges. Such divergence does not occur in our Hamiltonian model as the two-particle potential remains finite due the twist between adjacent base pairs. Nevertheless, a truncation of the configuration space for the base pair amplitudes is required in the computation and the choice of the cutoffs carries some arbitrariness.
These issues are handled in the computational method based on the path integral formalism as outlined in Section IV.

\section*{IV. Computational Method}

In our method, the base pair separations $r_i$  are thought of as trajectories $r_i(\tau)$ depending on the imaginary time $\tau=\,it$,  with $t$ being the real time for the evolution amplitude in the interval, $t_b - t_a$ \cite{fehi,jack}. Hence, $\tau$ varies in a range $\tau_b - \tau_a$ whose amplitude is set by the inverse temperature $\beta$ and the partition function for the Hamiltonian in Eq.~(\ref{eq:01}) is obtained by integrating over closed trajectories, $(\,r_i(0)=\, r_i(\beta) \,)$,  defined along the $\tau$-axis, following a route extensively described in previous works \cite{io14b} .

Accordingly, $r_i(\tau)$ can be expanded in  Fourier series :

\begin{eqnarray}
& &r_i(\tau)=\, (r_0)_i + \sum_{m=1}^{\infty}\Bigl[(a_m)_i \cos(\omega_m \tau ) + (b_m)_i \sin(\omega_m \tau ) \Bigr] \, , \nonumber
\\
& &\omega_m =\, \frac{2 m \pi}{\beta} 
\label{eq:03}
\end{eqnarray}

whose coefficients generate an ensemble of possible base pair paths and define the associated integration measure $\oint {D}r_i$:

\begin{eqnarray}
& &\oint {D}r_{i} \equiv  \frac{1}{\sqrt{2}\lambda_{cl}} \int_{-\Lambda^0_T}^{\Lambda^0_T} d (r_0)_i  \prod_{m=1}^{\infty}\Bigl( \frac{m \pi}{\lambda_{cl}} \Bigr)^2 \int_{-\Lambda_T}^{\Lambda_T} d(a_m)_i \int_{-\Lambda_T}^{\Lambda_T} d(b_m)_i \, , 
\label{eq:04}
\end{eqnarray}

where $\Lambda^0_T$ and $\Lambda_T$ are the temperature dependent cutoffs and $\lambda_{cl}$ is the classical thermal wavelength \cite{io11a}.

As Eq.~(\ref{eq:04}) normalizes the kinetic term in the action, i.e. \cite{io03}:

\begin{eqnarray}
\oint {D}r_i \exp\Bigl[- \int_0^\beta d\tau {\mu \over 2}\dot{r}_i(\tau)^2  \Bigr] = \,1 \, ,
\label{eq:05} \,
\end{eqnarray}

our formalism provides a consistent method to derive the cutoffs in the integration over the path configuration space thus avoiding the above mentioned arbitrariness usually encountered in the application of transfer integral techniques to mesoscopic Hamiltonian models. Furthermore, from Eq.~(\ref{eq:03}),~(\ref{eq:05}), one easily finds that $\Lambda^0_T$ and $\Lambda_T$ are $\propto \sqrt{T}$ hence, the maximum path amplitudes included in the partition function are larger at higher $T$ as expected on physical grounds.
Also note that Eq.~(\ref{eq:05}) sets the free energy zero for our system and it is fulfilled for any $\mu$ in accordance with the fact that the free energy of a classical system does not depend on $\mu$.

Writing Eq.~(\ref{eq:01}) in terms of the Fourier expansion in Eq.~(\ref{eq:03}) and summing over the bending and twisting degrees of freedom, we finally express the partition function $Z_N$ for the helical molecule with $N$ base pairs: 

\begin{eqnarray}
& &Z_N=\, \oint Dr_{1} \exp \bigl[- A_a[r_1] \bigr]   \prod_{i=2}^{N}  \int_{- \phi_{M} }^{\phi_{M} } d \phi_i \int_{- \theta_{M} }^{\theta _{M} } d \theta_{i} \oint Dr_{i}  \exp \bigl[- A_b [r_i, r_{i-1}, \phi_i, \theta_i] \bigr] \, , \nonumber
\\
& &A_a[r_1]= \,  \int_{0}^{\beta} d\tau H_a[r_1(\tau)] \, , \nonumber
\\
& &A_b[r_i, r_{i-1}, \phi_i, \theta_i]= \,  \int_{0}^{\beta} d\tau H_b[r_i(\tau), r_{i-1}(\tau), \phi_i, \theta_i] \, ,
\label{eq:06}
\end{eqnarray}

where $\phi_{M}$ and $\theta_{M}$ are the cutoffs on the amplitudes of the bending and twisting fluctuations, respectively.

Then, our computational technique consists of a direct sum over a set of molecule configurations which statistically contribute to the path integral with a Boltzmann weight. Specifically, the ensemble size, given by the number of trajectories for any base pair in the chain, is enlarged until $Z_N$ numerically converges i.e., the state of thermodynamic equilibrium is achieved.  {This amounts to sum over $\sim 10^{8}$ configurations for each dimer in the chain.} Over such ensemble, we perform the averages to obtain the macroscopic helical parameters as described in the next Section.

While Eq.~(\ref{eq:06}) holds at any temperature, the Fourier expansion in Eq.~(\ref{eq:03}) is particularly useful at low and intermediate temperatures (up to room $T$) as it generates a large ensemble of paths in the configuration space. However, above room temperatures, considerable CPU time savings are enabled by taking only the zero mode in Eq.~(\ref{eq:03}), $r_i(\tau) \sim \, (r_0)_i$, so that the $d\tau$ integrals in Eq.~(\ref{eq:06}) are straightforward and the partition function reduces to: 

\begin{eqnarray}
& &Z_N \rightarrow \, Z_1  \cdot \prod_{i=2}^{N}  Z_i \, , \nonumber
\\
& &Z_1 =\, \frac{1}{\sqrt{2}\lambda_{cl}} \int_{-\Lambda^0_T}^{\Lambda^0_T} d(r_0)_{1} \exp \bigl[- A_a[(r_0)_1] \bigr] \, , \nonumber
\\
& &Z_i =\, \int_{- \phi_{M} }^{\phi_{M} } d \phi_i \int_{- \theta_{M} }^{\theta _{M} } d \theta_{i} \frac{1}{\sqrt{2}\lambda_{cl}} \int_{-\Lambda^0_T}^{\Lambda^0_T} d(r_0)_{i}  \exp \bigl[- A_b [(r_0)_i, (r_0)_{i-1}, \phi_i, \theta_i] \bigr] \, , \nonumber
\\
& &A_a[(r_0)_1]= \, \beta \cdot V_{1}[(r_0)_1] \, , \nonumber
\\
& &A_b[(r_0)_i, (r_0)_{i-1}, \phi_i, \theta_i]=\, \beta \cdot \Bigl( V_{1}[(r_0)_i] + V_{2}[ (r_0)_i, (r_0)_{i-1}, \phi_i, \theta_i] - F_{ex} d  \cos\bigl( \sum_{k=1}^{i-1}\phi_k \bigr) \Bigr)  \, . \nonumber
\\ 
\label{eq:07}
\end{eqnarray}

From Eq.~(\ref{eq:06}) or Eq.~(\ref{eq:07}), one derives the equilibrium thermodynamics from the free energy, $F=\, -\beta ^{-1} \ln Z_N$.  While the results hereafter presented are obtained from Eq.~(\ref{eq:06}), the same qualitative trend is found via computation of Eq.~(\ref{eq:07}).

\section*{V. Rise distance and helix radius}

As the focus of this work is on the helical torsional response to an applied load, we devise a recursive procedure to express the twist angle in Eq.~(\ref{eq:06}). 
Precisely, $\theta_{i}$ is measured with respect to the ensemble average $<\theta_{i - 1}>$ obtained for the preceding base pair along the molecule stack (see Fig.~\ref{fig:1}(b)) and the latter value is incremented by $2\pi / h$, where $h$, the number of base pairs per helix turn, is chosen within a physically suitable range. Moreover, for any $h$ in such range, we integrate over a twist fluctuation variable $\theta_{i}^{fl}$ around the value $\, <\theta_{i - 1}>  + 2\pi / h \,$.
This idea is formally expressed by:

\begin{eqnarray}
& &\theta_i =\, <\theta_{i - 1}>  + 2\pi / h + \theta_{i}^{fl} ,  \nonumber
\\
& &h \in \, [h_{min}, \, h_{max}]  ,  \nonumber
\\
& &h_{max} - h_{min}=\, n \cdot \Delta h \,  \, ,
\label{eq:08}
\end{eqnarray}

where $n$ is the number of values sampled in the $h$ range and $\Delta h$ is the incremental step. Following the experiments, which yield $h^{exp}=\, 10.4$ for kilo-base B-DNA in solution under physiological condition \cite{wang,duguet}, we explore a wide range ($h_{min}=\,6$,\, $h_{max}=\,14$) around $h^{exp}$ as, in principle, short DNA chains may have a twist conformation which significantly differs from the long chains. Moreover, the molecule twist number is expected to vary with the applied load and the computational scheme should hold for any $F_{ex}$.

Using Eq.~(\ref{eq:08}) the average twists are computed by:

\begin{eqnarray}
& &< \theta_i >_{(i \geq 2)} =\,  < \theta_{i-1} > + \frac{2\pi}{h}  + \frac{\int_{-\theta_{M}}^{\theta_{M}} d \theta_{i}^{fl} \cdot ( \theta_i^{fl} ) \int_{- \phi _{M}}^{\phi _{M}} d \phi_i  \oint Dr_{i} \exp \bigl[- A_b [r_i, r_{i-1}, \phi_i, \theta_i]  \bigr]}{ \int_{-\theta_{M}}^{ \theta_{M}} d \theta_{i}^{fl} \int_{- \phi _{M} }^{\phi _{M} } d \phi_i  \oint Dr_{i} \exp \bigl[- A_b [r_i, r_{i-1}, \phi_i, \theta_i]  \bigr]} \, , \nonumber
\\
\label{eq:09}
\end{eqnarray}

while, for the first base pair in the chain, $< \theta_1 >=\,0$.   Hence, from Eq.~(\ref{eq:09}), we derive the average helical repeat as:

\begin{eqnarray}
< h >=\,\frac{2\pi N}{< \theta_N >} \, .
\label{eq:10}
\end{eqnarray}

Technically, the program sets an increment $\Delta h$ and, for any input $h$, computes a set of values $\{ < h >_{j} \, ,  \, (j=1,...,n) \} $ which differ from the initially chosen $h$-values. For any value $< h >_{j}$, the corresponding free energy is calculated.  As discussed in detail in ref.\cite{io17},  the procedure is reiterated by taking a finer $\Delta h$ until the average twists and associated free energies converge, i.e., they do not further change by increasing $n$. By minimizing $F$, one finally selects the equilibrium average twist conformation, denoted in the following by $< h >_{j^{*}}$, in the presence of a load. By varying $F_{ex}$, the calculation yields the twist-stretch profile for a specific molecule.  

{Likewise, one evaluates the average bending angles between adjacent base pair planes, $< \phi_i >_{(i \geq 2)}$, which also define the average angle between the stacking bonds and are measured with respect to the force direction, as shown in Fig.~\ref{fig:1}(b). By definition, $< \phi_1 > =\,0$. }

The numerical convergence of the ensemble integrations should be also tested against the maximum twisting and bending fluctuations. 
For the twist angles, we find that the appropriate cutoff to stabilize the ensemble averages is, $\theta_{M}=\,\pi /4$, which allows for large torsions between adjacent base pairs. As twisting fluctuations may be sizeable also in the presence of $F_{ex}$, there is no physical reason to introduce a force dependent twisting cutoff. 
 
In the absence of external loads, the bending cutoff  $\phi_{M}=\,\pi /2$ is large enough to allow for the formation of kinks having the effect to reduce the bending energy between neighbor base pairs \cite{crick,zocchi13,menon13,kim14}. It is however plausible that, by increasing $F_{ex}$, the intra-strand bonds should stretch and the amplitude of the bending cutoff should shrink. Although, at this stage, there is not experimental information to quantitatively account for such effect, we have tested some ad hoc functions $\phi_{M}(F_{ex})$ and chosen,  { $\phi_{M}(F_{ex})=\, {\pi }[ 1 - (c \cdot  F_{ex})^z ] / 2$ \,  with tunable parameters $c\, , z$ } in the calculations hereafter presented.

Analogously to Eq.~(\ref{eq:09}), ensemble averages are carried out to compute the macroscopic helical parameters, i.e., 
the average distances between adjacent base pairs along the stack,

\begin{eqnarray}
& &< \overline{d_{i,i-1}} > =\,  \frac{\oint Dr_{i}  \int_{-\theta_{M}}^{\theta_{M}} d \theta_{i}^{fl}   \int_{ -\phi _{M}}^{\phi _{M}} d \phi_i  \cdot \overline{d_{i,i-1}} \exp \bigl[- A_b [r_i, r_{i-1}, \phi_i, \theta_i]  \bigr]}{ \int_{-\theta_{M}}^{\theta_{M}} d \theta_{i}^{fl} \int_{-\phi _{M} }^{\phi _{M} } d \phi_i  \oint Dr_{i}  \exp \bigl[- A_b [r_i, r_{i-1}, \phi_i, \theta_i]  \bigr]} \, , \nonumber
\\
& &< d >=\frac{1}{N-1} \sum _{i=2}^{N} < \overline{d_{i,i-1}} > \, . 
\label{eq:11}
\end{eqnarray}

and the average base pair radial fluctuations:

\begin{eqnarray}
& &< r_i > =\,  \frac{\oint Dr_{i} \cdot r_i \int_{-\theta_{M}}^{\theta_{M}} d \theta_{i}^{fl}   \int_{-\phi _{M}}^{\phi _{M}} d \phi_i   \exp \bigl[- A_b [r_i, r_{i-1}, \phi_i, \theta_i]  \bigr]}{ \int_{-\theta_{M}}^{\theta_{M}} d \theta_{i}^{fl} \int_{-\phi _{M} }^{\phi _{M} } d \phi_i  \oint Dr_{i}  \exp \bigl[- A_b [r_i, r_{i-1}, \phi_i, \theta_i]  \bigr]} \, , \nonumber
\\
& &< R >=\frac{1}{N} \sum _{i=1}^{N} < r_i > \, 
\label{eq:12}
\end{eqnarray}

Note that the first base pair radial fluctuation, $< r_1 >$, is calculated with the Boltzmann weight given by the action $A_a[r_1]$ in Eq.~(\ref{eq:06}). 

{Before displaying the results, we discuss a possible extension of our model to the over-stretching regime in which the applied forces, in addition to aligning the stacking bonds, are strong enough to produce a sizeable elongation of the same bonds up to the rupture point. This can be accomplished as follows. First, one should compute at zero load, the ensemble averages $< \overline{d_{i,i-1}} >_{F_{ex}=\,0}$ over the radial and angular fluctuations and then consider the variations with respect to such averages in the presence of the applied load. Hence, assuming that the force acts in the same way on all dimers (and these lie along the force direction), the fourth addendum in the last of Eq.~(\ref{eq:01}) should be replaced by the term
$F_{ex} \bigl(\overline{d_{i,i-1}} - < \overline{d_{i,i-1}} >_{F_{ex}=\,0} \bigr) $ which in turn should be statistically weighted according to Eq.~(\ref{eq:06}).  For the purpose of the present analysis, the Hamiltonian in Eq.~(\ref{eq:01}) provides however a reasonable and computationally more convenient theoretical scheme. }

\section*{VI. Results}

\begin{figure}
\includegraphics[height=8.0cm,width=8.0cm,angle=-90]{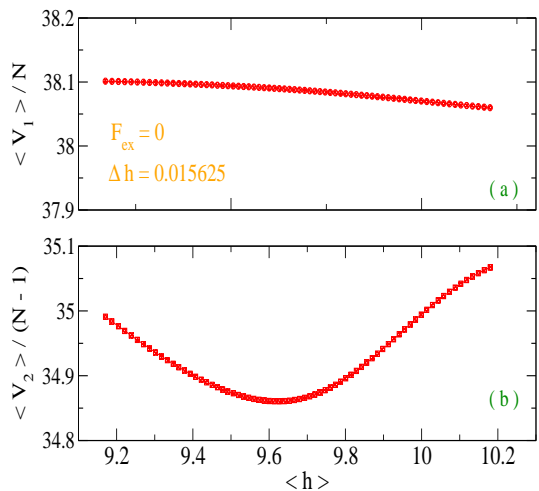}
\caption{\label{fig:2}(Color online) 
Ensemble averages of the (a) one-particle potential energy per base pair and (b) two-particle potential energy per dimer, defined in Eqs.~(\ref{eq:01}),~(\ref{eq:02}).
The average potential terms are plotted versus the average helical repeat calculated through Eqs.~(\ref{eq:09}),~(\ref{eq:10}) and are in units $meV$. 
}
\end{figure}

{The model is applied to a very short homogeneous fragment with $N=\,10$ base pairs. The calculation is carried out with a fine mesh, i.e. $\Delta h=\, 0.015625$ in Eq.~(\ref{eq:08}).  First, we compute in the absence of external loads, the ensemble averages of the one-particle and two-particle potential energy  of the Hamiltonian in Section III.  In Fig.~\ref{fig:2}, the terms $< V_1 > =\, < \sum_{i=1}^{N} V_{1}[r_i] >$ and   $< V_2 > =\, < \sum_{i=2}^{N} V_{2}[ r_i, r_{i-1}, \phi_i, \theta_i] >$  are plotted as a function of $< h >$.  While $< V_1 >$ shows a very weak dependence on $< h >$ ascribable to the interplay between radial and angular fluctuations only in the Boltzmann statistical weight,  $< V_2 >$ displays a pronounced minimum versus $< h >$ as the stacking explicitly depends on the twist angle. This points to the fact that some twist conformations are energetically favored.  }

\begin{figure}
\includegraphics[height=8.0cm,width=8.0cm,angle=-90]{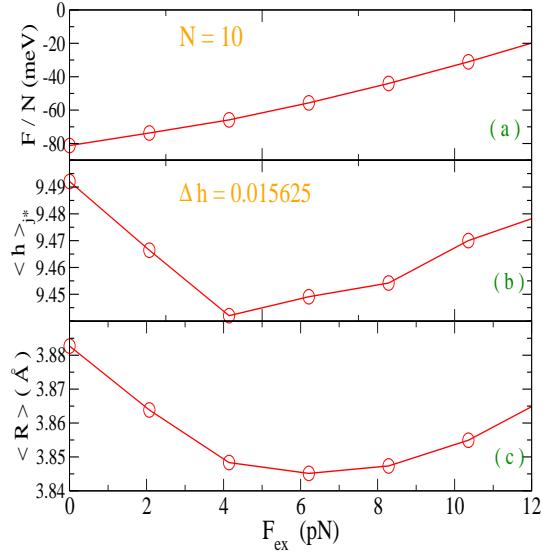}
\caption{\label{fig:3}(Color online) 
Thermodynamical properties and structural parameters of a homogeneous sequence with $N=\,10$ base pairs under a load $F_{ex}$. The bending cutoff function $\phi_{M}(F_{ex})$ is defined by the parameters, $c^{-1} = 24$pN and $z=\,2$. 
(a) Minimum free energy per base pair;
(b) average helical repeat determined by free energy minimization;
(c) average base pair radial fluctuation.
}
\end{figure}

{The precise evaluation of the equilibrium twist is carried out in  Fig.~\ref{fig:3} by minimizing the system free energy both for $F_{ex}=\,0$ and for a choice of loads applied to the fragment.}
For finite forces, the bending cutoff parameters are set as:\, $c^{-1}=\, 24$ pN and $z=\,2$. For any $F_{ex}$, the minimum free energy per base pair is plotted in the panel (a).

$F / N$ increases monotonically versus $F_{ex}$ consistently with the fact that external forces straighten the helix thus reducing the entropy of the chain.
These $F / N$ values correspond to the twist conformation specified by $< h >_{j^{*}}$ in the panel (b). By introducing a load, the short helix initially over-twists ($< h >_{j^{*}}$ decreases) and eventually untwists under stronger stretchings. The upturn is here found at $F_{ex} \sim 4 pN$,   the typical thermal energy per nano-meter mentioned in the Introduction, for the choice of $\phi_{M}(F_{ex})$ and model potential parameters given above. 
Physically, the over-twisting of the helix is accompanied by a contraction of the helix radius whereas the latter expands if the helix untwists. Indeed, this is the behavior displayed in the panel (c) where Eq.~(\ref{eq:12}) is calculated for the twist conformations given in the panel (b).

\begin{figure}
\includegraphics[height=8.0cm,width=8.0cm,angle=-90]{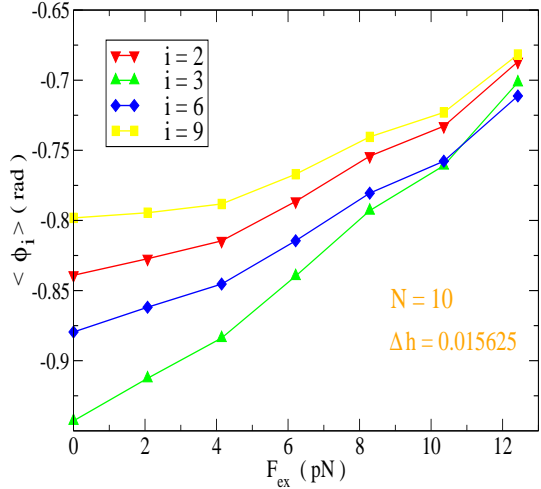}
\caption{\label{fig:4}(Color online) 
Ensemble averages of the bending angles between adjacent stacking bonds versus $F_{ex}$, for the same chain of Fig.~\ref{fig:3}. The average bending angles of four dimers are shown. The minus sign for the averages follows from the fact that the angles increase counterclockwise from the force direction, see Fig.~\ref{fig:1}(b).
}
\end{figure}

{ The role of the applied load is highlighted, at the level of the base pairs, in Fig.~\ref{fig:4} where the $< \phi_i >_{(i \geq 2)}$'s are plotted for the same force values taken in Fig.~\ref{fig:3}. Only four average angles are reported for clarity. At zero load, the $< \phi_i >_{}$'s are distributed over a range of values consistently with the coiled conformation of the molecule. By enhancing the load, such range markedly narrows and the absolute values of $< \phi_i >_{}$'s get smaller. This  corresponds physically to the progressive alignment of the stacking bonds to the force direction. }

\begin{figure}
\includegraphics[height=8.0cm,width=8.0cm,angle=-90]{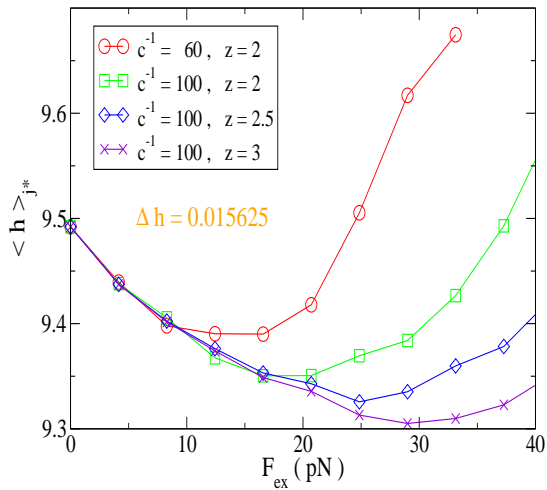}
\caption{\label{fig:5}(Color online) 
Average helical repeat of a homogeneous fragment with $N=\,10$ base pairs versus the external load $F_{ex}$. The computation is performed as in Fig.~\ref{fig:3} but different forms for the parameters $c,\, z$ of the bending cutoff $\phi_{M}(F_{ex})$ are assumed (see text).  $c^{-1}$ is in units $pN$.
}
\end{figure}

Experiments on kilo-base sequences have located the upturn at $F_{ex} \sim 30 pN$ \cite{busta06}.  Analogous experiments on fragments with a few tens of base pairs would help to check whether a similar twisting pattern persists at short length scales and also serve as a criterion to fit the parameters of our model. 
While in principle the behavior of kilo-base sequences may differ from that of very short fragments,  we have repeated the free energy minimization procedure for the same sequence of  Fig.~\ref{fig:3} varying the parameters of the bending cutoff $\phi_{M}(F_{ex})$.  
Some results are displayed in  Fig.~\ref{fig:5}.  The computation shows that the occurrence of the over-twisting / untwisting transition is indeed sensitive to the specific dependence of the maximum amplitude of the bending fluctuations on the applied load. By reducing $c$ and (or) increasing $z$ with respect to  Fig.~\ref{fig:3},  we assume that the maximum bending fluctuations decrease more smoothly for large loads and, under these conditions, the over-twisting regime progressively extends up to a few tens of picoNewtons. At last, it is found that for $c^{-1}=\,100$pN and $z=\,3$, the upturn in the $< h >_{j^{*}}$  plot can be set at $F_{ex} \sim 30 pN$ as in Ref.\cite{busta06}.  For the latter curve, we can estimate the superhelical density as a function of the load, $\sigma (F_{ex}) =\, \Delta Tw / (Tw)_{0}$, where $\Delta Tw$ is the number of turns added to the helix by increasing the stretching perturbation and $(Tw)_{0}$ is the unperturbed twist number. 
For instance, taking  the calculated $< h >_{j^{*}}$ at $F_{ex} \sim 8.3 pN$, we obtain $\sigma (F_{ex}) =\, 0.0098$.

\begin{figure}
\includegraphics[height=8.0cm,width=8.0cm,angle=-90]{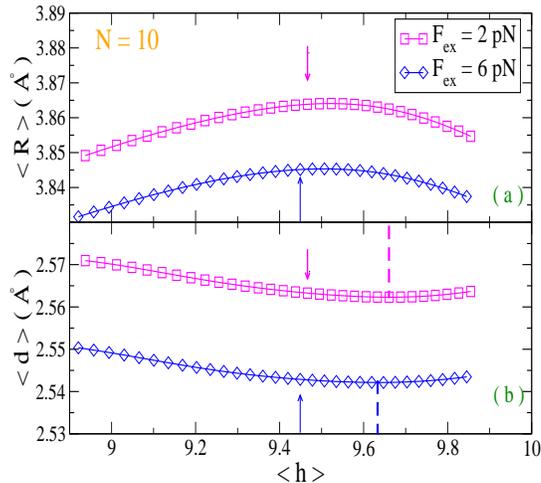}
\caption{\label{fig:6}(Color online)  
(a) Average base pair radial fluctuation and (b) average rise distance are computed, for the same chain and model parameters of Fig.~\ref{fig:3}, as a function of the average helical repeat.
Two external loads are considered.
The arrows mark the values corresponding to the equilibrium helical conformation $< h >_{j*}$ plotted in Fig.~\ref{fig:3}. For $< h >$ smaller than $< h >_{j*}$, the helix is over-twisted.
}
\end{figure}

Our code however, computes the helix structural parameters for the whole set of conformations obtained from Eqs.~(\ref{eq:08})-(\ref{eq:10}) and not only for the conformation $< h >_{j^{*}}$ (reported in Figs.~\ref{fig:2},~\ref{fig:3}) which minimizes the free energy. Then we can monitor how the helical shape changes, away from the minimum, assuming an over-twisted conformation ($< h >$ smaller than $< h >_{j^{*}}$) or an untwisted conformation ($< h >$ larger than $< h >_{j^{*}}$). As this is done at a fixed external load, our reasoning simulates the experimental setups of ref.\cite{busta06,croq06} in which small torsional strains are applied to the molecule under constant tension. Fig.~\ref{fig:6} shows our findings for the same molecule of Fig.~\ref{fig:3} and for two distinct loads. We see that an imposed over-twist causes the average helix diameter to shrink (panel (a)) and the average rise distance to elongate (panel (b)) with respect to the values corresponding to the equilibrium conformation $< h >_{j^{*}}$. Interestingly, if we slightly untwist the helix, the rise distance shortens  (down to the dashed lines in panel (b)) whereas $< d >$ eventually extends if we further untwist the helix (reaching the regime to the right of the dashed lines).  This trend holds for any constant load in qualitative agreement with the behavior of kilo-base sequences of Ref.\cite{busta06}.

\begin{figure}
\includegraphics[height=8.0cm,width=8.0cm,angle=-90]{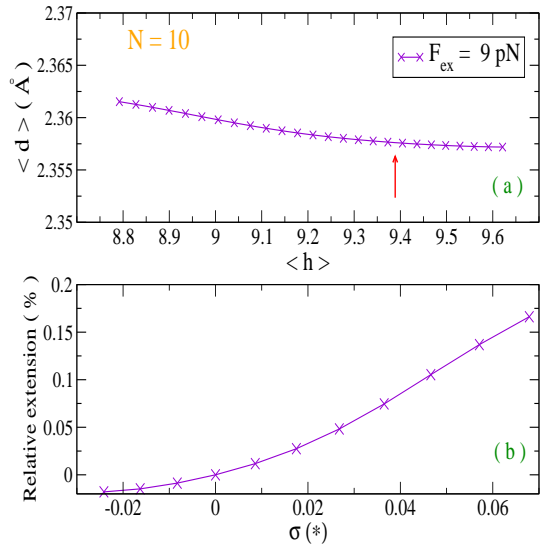}
\caption{\label{fig:7}(Color online)  
(a) Average rise distance versus average helical repeat for the same sequence of Fig.~\ref{fig:5}. The bending cutoff parameters are $c^{-1}=\, 100$pN,  $z=\,3$.
The arrow marks the value $< d >_{j*}$ corresponding to the equilibrium helical conformation $< h >_{j*}$, obtained in Fig.~\ref{fig:5}, with external load $F_{ex}=\,9 pN$. For $< h >$ smaller than $< h >_{j*}$, the helix is over-twisted.  For $< d >$ larger than $< d >_{j*}$, the average intra-strand base pair distance is stretched with respect to the equilibrium.  (b) Relative extension versus superhelical density ($\sigma (j*) \equiv (\Delta Tw)_{j*} / (Tw)_{j*}$) derived from (a).
}
\end{figure}

{Likewise, assuming the bending cutoff parameters which set the upturn at  $F_{ex} \sim 30 pN$ in Fig.~\ref{fig:5}, we find that $< d >$ for the fragment with $N=\,10$ base pairs grows by over-twisting the helix with respect to its equilibrium twist conformation, at constant force. The results are shown in Fig.~\ref{fig:7}(a) for  $F_{ex} \sim 9 pN$ as, for such load, some data are available from Ref.\cite{busta06}. Also the percent relative extension ($\equiv \, {< d >}/{< d >_{j*}} - 1$) is reported in Fig.~\ref{fig:7}(b) as a function of the superhelical density $\sigma (*)=\, (\Delta Tw) _{j*} / (Tw)_{j*}$. Note that $\sigma (*)$ is defined with respect to the equilibrium twist number ($(Tw)_{j*}=\, N / < h >_{j*}$)  and $(\Delta Tw) _{j*}$ measures the applied torsional strain \textit{at fixed force}. Hence, $\sigma (*)$ is physically distinct from the previously defined $\sigma (F_{ex})$.  The  intra-strand distance grows almost linearly above $\sigma (*) \sim 0.01$  but, at the latter value, we obtain a relative increment which is an order of magnitude lower than that reported in Ref.\cite{busta06}. It is however remarked that, for any given $\sigma (*)$, we are simulating $(\Delta Tw)_{j*}$ values which are too small to allow a strict comparison with data relative to kilo-base sequences.
}

\begin{figure}
\includegraphics[height=8.0cm,width=8.0cm,angle=-90]{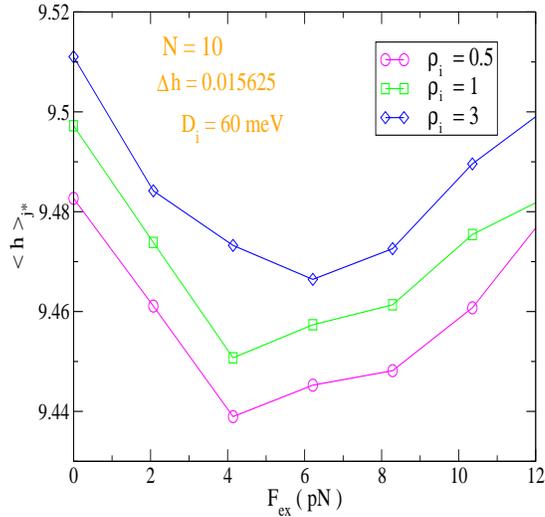}
\caption{\label{fig:8}(Color online)  
Average helical repeat, calculated by minimizing the free energy of the homogeneous fragment, versus the external force. The three plots are obtained by varying the nonlinear stacking parameter $\rho_i$ in  Eq.~(\ref{eq:02}).
}
\end{figure}

Finally, one may wonder to which extent the results presented so far are sensitive to the choice of the model potential parameters. In Fig.~\ref{fig:8}, we take the $N=\,10$ homogeneous sequence with base pair dissociation energy $D_i=\, 60 meV$ (higher than in Fig.~\ref{fig:3}) and plot the average helical repeat as function of the external load, for three values of the nonlinear stacking parameter $\rho_i$. All other model parameters are as in Fig.~\ref{fig:3}.  The incremental step, $\Delta h$ in Eq.~(\ref{eq:08}), is the same as in Fig.~\ref{fig:3} and the $< h >_{j^{*}}$ are computed by the free energy minimization method discussed above. 

For all $\rho_i$'s, we notice that the helix over-twists under small loads and then untwists at larger forces. However the equilibrium twist conformations markedly depend, at any $F_{ex}$, on the size of the nonlinear stacking: precisely, the calculation shows that larger $\rho_i$ values induce higher $< h >_{j^{*}}$ as expected on physical grounds. 
{In fact as discussed in Section III, when thermal fluctuations cause the transient breaking of a hydrogen bond, larger $\rho_i$'s increase the energetic gain associated to the unstacking of adjacent bases. Therefore larger $\rho_i$'s favor equilibrium conformations with a higher number of base pairs per helix turn. }
Also note that, for the chain with $\rho_i=\,3$, the force induced transition between over-twisting and untwisting regime is shifted at larger forces with respect to the chains with smaller $\rho_i$. As experiments on kilo-base DNA sequences have observed the transition at $F_{ex} \sim 30 pN$, our result may suggest that longer sequences have higher intrinsic flexibility (larger $\rho_i$ in our model) and therefore they are more resilient to the stretching perturbations. Hence, they begin to untwist only under sizeable loads.

Comparing the green plot ($\rho_i=\,1$) with Fig.~\ref{fig:3}(b),  it is found that the $< h >_{j^{*}}$'s values become slightly larger by increasing $D_i$. This result is apparently surprising as one may expect that higher hydrogen bond energies stabilize the molecule and oppose the helix untwisting. In fact, assuming higher $D_i$'s with all other parameters unchanged, yields more tightly bound complementary strands ($< R > $ shrinks) whereas the intra-strand rise distance $< d > $  grows (not shown here). As a consequence the molecule of the green plot in Fig.~\ref{fig:8} appears, with respect to the molecule in Fig.~\ref{fig:3}, in a narrower albeit more elongated configuration which ultimately favors a slight untwisting and higher $< h >_{j^{*}}$'s.  {Note however that the pair dissociation energies are taken within a range consistent with the experimental free energies per base pair } \cite{io16b}. These findings have been qualitatively discussed to point out that, by virtue of the interplay between inter-strand and intra-strand base pair interactions, the model potential parameters are intertwined and should be determined, as a set, by fitting the available data for specific fragments \cite{weber09}.

\section*{VII. Conclusions }

The mechanical response of DNA molecules to applied loads provides insights into the intrinsic flexibility properties of the helix which may vary with its sequence specificity, length and environmental conditions. While the theory of elastic rods generally accounts for the behavior of kilo-base pair filaments whose contour length  largely exceeds their characteristic persistence length, current research is revealing that, at short length scales, all-atom simulations and mesoscopic models should be rather used to capture the elastic behavior of fragments which display strong base pair fluctuational effects. Following our previous investigations on the cyclization properties, end-to-end distance and persistence length of sequences with $100$ base pairs or less, we have here studied how the helix macroscopic parameters and its twisting conformation may change under the effect of an external load. The computational method is based on a path integral description for the Hamiltonian model which comprises both inter-strand hydrogen bonds between the pair mates and intra-strand forces between adjacent bases along the molecule stack. Base pair separations are treated as trajectories contributing to the partition function with their specific statistical weight which is essentially determined by the physical constraints of the model potential. Thus, too large contractions of the base pair distance (with respect to the equilibrium helix diameter) are energetically discouraged by the electrostatic repulsions and therefore yield scarce weight to the partition function. On the other hand, the maximum amplitude of the base pair separations encounters in the path integration a temperature dependent cutoff which consistently truncates the configuration space. Crucial to our study are the bending and twisting fluctuations between neighbor base pairs which characterize the form of the stacking potential and also contribute to the partition function. Technically, the density of the base pair paths and angular fluctuations included in the computation is increased until the molecule free energy converges. For any applied force in a suitable pico-Newton range, it is assumed that the molecule may exist in a large  set of possible twist conformations each characterized by an average number of base pairs per helix turn. Carrying out the ensemble averages over the base pair degrees of freedom, we eventually obtain the average equilibrium helical twist by minimizing the system free energy. Moreover, we compute the helical twist, away from the equilibrium, associated to over-twisted and un-twisted conformations which can be simulated keeping the external load constant. Thus, our numerical program can predict:\textit{ a)} how the helical twist changes by varying the load and \textit{b)} how the helical parameters, i.e. average diameter and rise distance, change by over-twisting (untwisting) the helix at constant load. 

Applying the method to a very short homogeneous helix, we have found a general pattern similar to that  observed in kilo-base long sequences although short fragments may present macroscopic helix parameters which quantitatively differ from the long ones. 

Specifically, tuning the cutoff in the bending fluctuations integration, the model can even reproduce the transition between over-twisting and untwisting regime experimentally located at $\sim 30 pN$ in kilo-base sequences, with the caveat that such forces may be excessive to be applied to fragments of only ten base pairs  { as they could disrupt the stacking bonds.} Furthermore, adding (and subtracting) helix turns to the equilibrium conformation for a fixed stretching perturbation, we have derived the relation between helix elongation and superhelical density which appears essentially linear in the over-twisting regime.

We have also shown how the equilibrium helical repeat, predicted by minimizing the free energy in the presence of a load, could vary  
with the specific choice of model potential parametrization. Thus, we feel confident that the proposed method makes a valid computational tool to characterize specific fragments for which experiments could provide sufficient information to fit the set of potential parameters.


\begin{thebibliography}{widest-label}

\bibitem{stasiak}
K. Kiianitsa, A. Stasiak, Helical repeat of DNA in the region of homologous pairing. \emph{Proc. Natl. Acad. Sci. USA},  \textbf{94}, 7837-7840, 1997.

\bibitem{forget}
A.L. Forget, S.C. Kowalczykowski,  Single-molecule imaging of DNA pairing by RecA reveals a three-dimensional homology search. \textit{Nature},  \textbf{482}, 423, 2012. 

\bibitem{cappello} 
S. Atwell, L. Disseau, A.Z. Stasiak, A. Stasiak,
A. Renodon-Corni\`{e}re, M. Takahashi, J.-L. Viovy and G. Cappello,  Probing Rad51-DNA interactions by changing DNA twist.  \emph{Nucleic Acids Res.},  \textbf{40}, 11769-11776, 2012.

\bibitem{wang12}
J.L. Killian, M. Li, M.Y. Sheinin, M.D. Wang,  Recent advances in single molecule studies of nucleosomes. \textit{Curr. Opin. Struct. Biol.},   \textbf{22}, 80–87, 2012.

\bibitem{albu14a}
E.L. Albuquerque, U.L. Fulco, V.N. Freire, E.W.S. Caetano, M.L. Lyra, and F.A.B.F. de Moura, DNA-Based Nanobiostructured Devices:
The Role of the Quasiperiodicity and Correlation Effects. \textit{Phys. Rep.}, \textbf{535}, 
139, 2014.

\bibitem{chu}
S. Chu,   Laser manipulation of atoms and particles. \emph{Science},   \textbf{253},  861-866,  1991.

\bibitem{busta92}
S. Smith, L. Finzi, C. Bustamante, Direct mechanical measurement of the elasticity of single DNA molecules by using magnetic beads.  \emph{Science},   \textbf{258},  1122-1126, 1992.

\bibitem{busta94}
C. Bustamante, J.F. Marko, E.D. Siggia, S. Smith,   Entropic elasticity of lambda-phage DNA. \emph{Science},   \textbf{265},  1599-1601, 1994.

\bibitem{cluzel96}
P. Cluzel, A. Lebrun, C. Heller, R. Lavery, J.L. Viovy, D. Chatenay, F. Caron,   DNA: an extensible molecule. \emph{Science},   \textbf{271},  792-794, 1996.

\bibitem{block97} 
M.D. Wang, H. Yin, R. Landick, J. Gelles, S.M. Block,  Stretching DNA with Optical Tweezers.  \textit{Biophys. J.},    \textbf{72}, 1335-1346,  1997.

\bibitem{mameren09} 
J. van Mameren, P. Gross, G. Farge, P. Hooijman, M. Modesti, M. Falkenberg,
G.J.L. Wuite and E.J.G. Peterman,   Unraveling the structure of DNA during overstretching by using multicolor, single-molecule fluorescence imaging. 
\emph{Proc. Natl. Acad. Sci. USA},  \textbf{106},  18231-18236, 2009.

\bibitem{busta00}
C. Bustamante, S.B. Smith, J. Liphardt, D. Smith,   Single-molecule studies of DNA mechanics.  \textit{Curr. Opin. Struct. Biol.},  \textbf{10}, 279-285, 2000.

\bibitem{rouz01}
I. Rouzina, V.A. Bloomfield,  Force-Induced Melting of the DNA Double Helix 1. Thermodynamic Analysis.  \emph{Biophys. J.},   \textbf{80}, 882-893, 2001.

\bibitem{busta06} 
J. Gore, Z. Bryant, M. N\"{o}llmann, M.U. Le, N.R. Cozzarelli and C. Bustamante,   DNA overwinds when stretched. \textit{ Nature},  \textbf{442}, 836-839, 2006.


\bibitem{croq06}
T. Lionnet,  S. Joubaud,  R. Lavery,  D. Bensimon, V. Croquette,   Wringing out DNA.  \emph{Phys. Rev. Lett.},  \textbf{96}, 178102, 2006.

\bibitem{zachar15}
K. Liebl, T. Drsata, F. Lankas, J. Lipfert,  M. Zacharias,  Explaining the striking difference in twist-stretch
coupling between DNA and RNA: A comparative molecular dynamics analysis. \textit{Nucleic Acid Res.},  \textbf{43},  10143-10156, 2015. 

\bibitem{marko97}
J.F. Marko,  Stretching must twist DNA.  \emph{Europhys. Lett.},   {\bf 38}, 183-188, 1997. 

\bibitem{wuite11} 
P. Gross, N. Laurens, L.B. Oddershede, U. Bockelmann, E.J.G. Peterman and G.J.L. Wuite,  Quantifying how DNA stretches, melts and changes twist under tension.   \textit{ Nature Phys.},  \textbf{7}, 731–736, 2011.

\bibitem{io17}
M. Zoli,  Twist-stretch profiles of DNA chains. \textit{J. Phys.: Condens. Matter},  \textbf{29},  225101, 2017. 

\bibitem{campa}
A. Campa, A. Giansanti,  Experimental tests of the Peyrard-Bishop model applied to the melting of very short DNA chains.   \textit{Phys. Rev. E},  \textbf{58}, 3585-3588,   1998.

\bibitem{albu14}
D.X. Macedo, I. Guedes, E.L. Albuquerque,  Thermal properties of a DNA denaturation with
solvent interaction.  \textit{Physica A},  \textbf{404}, 234-241, 2014.

\bibitem{weber15a}
G. Weber,  Optimization method for obtaining nearest-neighbour DNA entropies and enthalpies directly from melting temperatures.  \textit{Bioinformatics} \textbf{31}, 871–877, 2015.

\bibitem{singh15}
A. Singh, N. Singh,  Effect of salt concentration on the stability of heterogeneous DNA.   \textit{Physica A},  \textbf{419}, 328-334, 2015.

\bibitem{widom}
T.E. Cloutier, J. Widom,    Spontaneous Sharp Bending of Double-Stranded DNA.  \emph{Mol. Cell.},  \textbf{ 14}, 355-362,  2004.

\bibitem{volo05}
Q. Du,  C. Smith, N. Shiffeldrim, M. Vologodskaia, and A. Vologodskii,  Cyclization of short DNA fragments and bending fluctuations of the double helix.   \emph{Proc. Natl. Acad. Sci. USA}, \textbf{102},  5397-5402,   2005.

\bibitem{olson06}
L. Czapla, D. Swigon, W.K. Olson,   Sequence-dependent effects in the cyclization of short DNA.  \emph{J. Chem. Theory Comput.}, \textbf{2},  685-695, 2006.

\bibitem{archer} 
C. Yuan,  H. Chen, X.W. Lou,  L.A. Archer,   DNA bending stiffness on small length scales.  \textit{Phys. Rev. Lett.},  {\bf 100}, 018102,  2008.

\bibitem{fenn}
R.S. Mathew-Fenn, R. Das,  P.A B. Harbury,   Remeasuring the Double Helix. \textit{Science}, \textbf{322}, 446-449,  2008.

\bibitem{mastro} 
A.J. Mastroianni, D.A. Sivak, P.L. Geissler, A.P. Alivisatos,  Probing the Conformational Distributions of Subpersistence Length DNA.  \emph{Biophys. J.},   \textbf{97},  1408-1417, 2009.


\bibitem{gole} 
A. Noy, R. Golestanian,  Length Scale Dependence of DNA Mechanical Properties. \textit{Phys. Rev. Lett.},   {\bf 109}, 228101,  2012.

\bibitem{vafa}
R. Vafabakhsh, T. Ha,  Extreme Bendability of DNA Less than 100 Base Pairs Long Revealed by Single-Molecule Cyclization.  \emph{Science},  \textbf{337}, 1097-1101,  2012.

\bibitem{kim14}
T.T. Le, H.D. Kim,    Probing the elastic limit of DNA bending.  \emph{Nucleic Acids Res.},   \textbf{42}, 10786-10794,  2014.

\bibitem{mazur14}
A.K. Mazur, M. Maaloum,  DNA Flexibility on Short Length Scales Probed by Atomic Force Microscopy.  \emph{Phys. Rev. Lett.},   \textbf{112},  068104,  2014. 

\bibitem{tan15} 
Y.Y. Wu, L. Bao, X. Zhang, Z.J. Tan,    Flexibility of short DNA helices with finite-length effect: from base pairs to tens of base pairs.  \emph{J. Chem. Phys.},  \textbf{142}, 125103, 2015.

\bibitem{io16a} 
M. Zoli,  Flexibility of short DNA helices under mechanical stretching. \textit{Phys. Chem. Chem. Phys.},  {\bf 18}, 17666, 2016.  

\bibitem{zhen17}
P.-M. Lam, Y. Zhen,  Cyclization of short DNA fragments. \textit{Physica A},  \textbf{482}, 569–572, 2017.

\bibitem{tan16}
L. Bao, X. Zhang, L. Jin,  Z.-J. Tan, Flexibility of nucleic acids: From DNA to RNA. \textit{Chin. Phys. B},  \textbf{25},   018703, 2016.

\bibitem{volo10}
S. Geggier, A. Vologodskii,   Sequence dependence of DNA bending rigidity.  \emph{Proc. Natl. Acad. Sci. USA},   \textbf{107}, 15421-15426, 2010.

\bibitem{olson10}
G. Zheng, L. Czapla, A. R. Srinivasan, W.K. Olson, How stiff is DNA?   \emph{Phys. Chem. Chem. Phys. },   \textbf{12},  1399–1406,  2010.

\bibitem{piana05}
S. Piana, Structure and energy of a DNA dodecamer under tensile load. \emph{Nucleic Acids Res.}, \textbf{33}, 7029-7038, 2005.

\bibitem{pey04}
M. Peyrard,    Nonlinear dynamics and statistical physics of DNA.   \textit{Nonlinearity},  {\bf 17}, R1,   2004.

\bibitem{io09}
M. Zoli,   Path integral method for DNA denaturation. \emph{Phys.Rev. E},   \textbf{79},  041927, 2009.

\bibitem{io10}
M. Zoli,  Denaturation patterns in heterogeneous DNA. \emph{Phys.Rev. E},    \textbf{81},  051910, 2010. 

\bibitem{tan08}
Z.J. Tan, S.J. Chen,  Electrostatic free energy landscapes for DNA helix bending. \textit{Biophys. J.},  \textbf{94}, 3137-3149,  2008.

\bibitem{orozco15}
G. Rossetti, P.D. Dans, I. Gomez-Pinto, I. Ivani, C. Gonzalez, M. Orozco,  The structural impact of DNA mismatches. \textit{Nucleic Acid Res.},  \textbf{43},  4309–4321, 2015.

\bibitem{hwang16}
X. Teng,  W. Hwang,  Elastic Energy Partitioning in DNA Deformation and Binding to Proteins. \textit{ACS Nano}, \textbf{10}, 170-180, 2016.

\bibitem{olson01} 
W.K. Olson, M. Bansal, S.K. Burley, R.E. Dickerson, M. Gerstein, S.C. Harvey,
U. Heinemann, X.-J. Lu, S. Neidle, Z. Shakked, H. Sklenar, M. Suzuki, C.-S. Tung, E. Westhof,
C. Wolberger and H.M. Berman,  A Standard Reference Frame for the Description of
Nucleic Acid Base-pair Geometry. \emph{J. Mol. Biol.}, \textbf{313}, 229-237, 2001.

\bibitem{lavery10}  
R. Lavery, K. Zakrzewska, D. Beveridge, T.C. Bishop, D.A. Case, T. Cheatham III, S. Dixit, B. Jayaram, F. Lankas,
C. Laughton, J.H. Maddocks, A. Michon, R. Osman, M. Orozco, A. Perez, T. Singh, N. Spackova and J. Sponer,  A systematic molecular dynamics study of nearest-neighbor effects on base pair and base pair step conformations and fluctuations in B-DNA. \emph{Nucleic Acids Res.}, \textbf{38}, 299-313, 2010.


\bibitem{bates}
A.D. Bates, A. Maxwell,   \emph{DNA Topology} (Oxford University Press, Oxford), 2009.


\bibitem{irob}
R.N. Irobalieva, J.M. Fogg, D.J. Catanese, T. Sutthibutpong, M. Chen, A.K. Barker, S.J. Ludtke, S.A. Harris, M.F. Schmid,
W. Chiu and L. Zechiedrich,     Structural diversity of supercoiled DNA.  \textit{Nat. Commun.},  \textbf{6}, 8440,   2015.

\bibitem{shore}
D. Shore, R.L. Baldwin,  Energetics of DNA twisting.  \emph{J. Mol. Biol.},  \textbf{170}, 983-1007, 1983. 

\bibitem{horowitz}
D. Horowitz, J. Wang,  Torsional Rigidity of DNA and Length Dependence of the Free Energy of DNA Supercoiling. \emph{J. Mol. Biol.}, \textbf{173}, 75-91, 1984.

\bibitem{seno10}
D. Marenduzzo,  E. Orlandini,  F. Seno,  A. Trovato,  Different pulling modes in DNA overstretching: A theoretical analysis. \textit{Phys. Rev. E}, \textbf{81}, 051926, 2010. 

\bibitem{romano13} 
F. Romano,  D. Chakraborty,  J.P.K. Doye, T.E. Ouldridge,  A.A. Louis, Coarse-grained simulations of DNA overstretching.  \emph{J. Chem. Phys.},   \textbf{138}, 085101, 2013.

\bibitem{fiasconaro13} 
A.E.B. Pupo,  F. Falo,  A. Fiasconaro, DNA overstretching transition induced by melting in a dynamical
mesoscopic model. \emph{J. Chem. Phys.},   \textbf{139}, 095101, 2013.

\bibitem{gaub08}
C.H. Albrecht, G. Neuert,  R.A. Lugmaier, H.E. Gaub, Molecular Force Balance Measurements Reveal that Double-Stranded
DNA Unbinds Under Force in Rate-Dependent Pathways. \emph{Biophys. J.},  \textbf{94}, 4766-4774, 2008.


\bibitem{wart}
R.M. Wartell, A.S. Benight,  Thermal denaturation of DNA molecules: A comparison of theory with experiment.  \emph{Phys. Rep.}, \textbf{126},  67-107,   1985.

\bibitem{zhang}
Y.L. Zhang, W.M. Zheng, J.X. Liu,  Y.Z. Chen,   Theory of DNA melting based on the Peyrard-Bishop model.  \textit{ Phys. Rev. E},  \textbf{56}, 7100-7115, 1997.


\bibitem{benham99}
R.M. Fye,  C.J. Benham,   Exact method for numerically analyzing a model of local denaturation
in superhelically stressed DNA.  \textit{Phys. Rev. E},  {\bf 59}, 3408-3426, 1999.

\bibitem{joy05}
M. Joyeux, S. Buyukdagli,    Dynamical model based on finite stacking enthalpies for homogeneous and inhomogeneous DNA
thermal denaturation.  \emph{Phys. Rev. E} {\bf 72},  051902, 2005.

\bibitem{zdrav06}
S. Zdravkovi\'{c}, M.V. Satari\'{c},   Single-molecule unzippering experiments on DNA and Peyrard-Bishop-Dauxois model. \emph{Phys. Rev. E},  {\bf 73},  021905, 2006.

\bibitem{krueg}
A. Krueger, E. Protozanova, M.D. Frank-Kamenetskii,   Sequence-Dependent Basepair Opening in DNA Double Helix. \textit{Biophys. J.},  {\bf 90}, 3091-3099, 2006.


\bibitem{kalos09}
G. Kalosakas, S. Ares,    Dependence on temperature and guanine-cytosine content
of bubble length distributions in DNA.   \textit{J. Chem. Phys.},   {\bf 130}, 235104, 2009.

\bibitem{pey09}
M. Peyrard, S. Cuesta-L\'{o}pez, D. Angelov,  Experimental and theoretical studies of
sequence effects on the fluctuation and melting of short DNA molecules.     \emph{J. Phys.: Condens. Matter}, \textbf{21}, 034103, 2009.

\bibitem{rapti10}
M.R. Kantorovitz, Z. Rapti, V. Gelev, A. Usheva,   Computing DNA duplex instability profiles
efficiently with a two-state model: trends of promoters and binding sites.  \textit{Bioinformatics},  \textbf{11}, 604, 2010.

\bibitem{bishop14}
C. Nisoli, A.R. Bishop,    Thermomechanical stability and mechanochemical response of DNA:
a minimal mesoscale model.  \emph{J. Chem. Phys.}, \textbf{141}, 115101, 2014.


\bibitem{weber15}
I. Ferreira, T.D. Amarante,  G. Weber,    DNA terminal base pairs have weaker hydrogen bonds especially for AT under low
salt concentration.  \emph{J. Chem. Phys.}, \textbf{143}, 175101, 2015.

\bibitem{singh16}
A. Singh, T. Modi, N. Singh,  Opening of DNA chain due to force applied on different locations.  \emph{Phys. Rev. E},  \textbf{94},   032410,   2016.


\bibitem{bonnet03}
G. Altan-Bonnet,  A. Libchaber, O. Krichevsky,   Bubble Dynamics in Double-Stranded DNA.   \emph{Phys. Rev. Lett.},  \textbf{90}, 138101, 2003.

\bibitem{zocchi04}
Y. Zeng, A. Montrichok, G. Zocchi,    Bubble Nucleation and Cooperativity in DNA Melting.  \emph{J. Mol. Biol.}, \textbf{339},  67-75, 2004.

\bibitem{rapti06}
Z. Rapti, A. Smerzi, K.{\O}. Rasmussen, A.R. Bishop, C.H. Choi, and A. Usheva, 
Healing length and bubble formation in DNA. \emph{Phys. Rev. E},   \textbf{73},   051902, 2006.

\bibitem{kalos11} 
A. Apostolaki, G. Kalosakas,   Targets of DNA-binding proteins in bacterial promoter regions present
enhanced probabilities for spontaneous thermal openings.  \textit{Phys. Biol.},  \textbf{8}, 026006, 2011. 

\bibitem{hand12}
A. Sulaiman, F.P. Zen,  H. Alatas,  L.T. Handoko,  The thermal denaturation of the
Peyrard-Bishop model with an external potential.  \textit{Phys. Scripta}, \textbf{86}, 015802,  2012. 

\bibitem{metz12}
J. Adamcik,  J.-H. Jeon,   K.J. Karczewski,  R. Metzler,  G. Dietler,   Quantifying supercoiling-induced denaturation bubbles in DNA.
\textit{Soft Matter},  \textbf{8}, 8651-8658, 2012.

\bibitem{palm13}
A.K. Dasanna, N. Destainville, J. Palmeri, M. Manghi,   Slow closure of denaturation bubbles in DNA: Twist matters.  \textit{Phys. Rev. E},  \textbf{87}, 052703, 2013.


\bibitem{poland}
D. Poland, H. Scheraga,   Occurrence of a Phase Transition in Nucleic Acid Models.  \emph{J. Chem. Phys.}, \textbf{45},  1464-1469,   1966.

\bibitem{proho85}
Y. Kim, K.V. Devi-Prasad, E.W. Prohofski,   Self-consistent phonon theory of mean-field hydrogen-bond melting of poly(DG)-poly(DC). \emph{Phys. Rev. B},  \textbf{32}, 5185-5189,  1985.

\bibitem{pey89}
M. Peyrard, A.R. Bishop,   Statistical mechanics of a nonlinear model for DNA denaturation. \emph{Phys. Rev. Lett.}, \textbf{62},  2755-2758, 1989.


\bibitem{io16b}
M. Zoli,   J- factors of short DNA molecules.  \textit{J. Chem. Phys.},   {\bf 144},  214104, 2016. 


\bibitem{druk}
K. Drukker, G. Wu, G.C. Schatz,   Model simulations of DNA denaturation dynamics. \emph{J. Chem. Phys.}, \textbf{114},  579-590,  2001.

\bibitem{io11}
M. Zoli,     Thermodynamics of twisted DNA with solvent interaction.  \emph{J. Chem. Phys.},  \textbf{135},  115101,  2011.

\bibitem{metz11}
S. Talukder, P. Chaudhury, R. Metzler, S.K. Banik,   Determining the DNA stability parameters for the breathing dynamics
of heterogeneous DNA by stochastic optimization.  \emph{J. Chem. Phys.},   \textbf{135},  165103, 2011.

\bibitem{dekker14}
J. Lipfert, G.M. Skinner, J.M. Keegstra, T. Hensgens, T. Jager, D. Dulin,
M. K\"{o}ber,  Z. Yu, S.P. Donkers, F.-C. Chou, R. Das,  N.H. Dekker, Double-stranded RNA under force and torque:
Similarities to and striking differences from double-stranded DNA. \emph{Proc. Natl. Acad. Sci. USA},   \textbf{111}, 15408-15413, 2014.


\bibitem{tan17} 
L. Bao, X. Zhang, Y.-Z. Shi, Y.Y. Wu, Z.J. Tan,   Understanding the Relative Flexibility of RNA and DNA Duplexes: Stretching and Twist-Stretch
Coupling. \emph{ Biophys. J.},  \textbf{112}, 1094-1104, 2017.


\bibitem{io12}
M. Zoli,    Anharmonic stacking in supercoiled DNA. \textit{J. Phys.: Condens. Matter},    {\bf 24},  195103, 2012.

\bibitem{pey2}
T. Dauxois, M. Peyrard, A.R. Bishop,   Entropy-driven DNA denaturation.  \emph{Phys. Rev. E},   \textbf{47},  R44-47, 1993.


\bibitem{ares}
S. Ares, N.K. Voulgarakis, K.{\O}. Rasmussen,  A.R. Bishop,   Bubble Nucleation and Cooperativity in DNA Melting.  \emph{Phys. Rev. Lett.}, \textbf{94},   035504,  2005.

\bibitem{fehi}
R.P. Feynman,  A.R. Hibbs,    {\it Quantum Mechanics and Path Integrals}, (Mc Graw-Hill, New York), 1965.

\bibitem{jack}
R. Jackiw,   Quantum meaning of classical field theory.  \textit{Rev. Mod. Phys.},  \textbf{49}, 681-706, 1977.

\bibitem{io11a}
M. Zoli,    Stacking interactions in denaturation of DNA fragments.  \textit{Eur. Phys. J. E},   {\bf 34}, 68, 2011.

\bibitem{io03}
M. Zoli,  Path integral description of a semiclassical Su-Schrieffer-Heeger model.  \textit{Phys. Rev. B},  {\bf 67}, 195102,  2003.

\bibitem{io14b} 
M. Zoli,   Twist versus nonlinear stacking in short DNA molecules.  \textit{J. Theor. Biol.},   {\bf 354},  95-104, 2014. 

\bibitem{wang}
J.C. Wang,   Helical repeat of DNA in solution.  \emph{Proc. Natl. Acad. Sci. USA},   \textbf{76},  200-203, 1979.

\bibitem{duguet}
M. Duguet,   The helical repeat of DNA at high temperature. \emph{Nucleic Acids Res.},  \textbf{21}, 463-468, 1993.


\bibitem{crick}
F.H. Crick, A. Klug,   Kinky helix.  \emph{Nature},  \textbf{255}, 530-533,  1975.

\bibitem{zocchi13}
D.S. Sanchez, H. Qu, D. Bulla, G. Zocchi,   DNA kinks and bubbles: Temperature dependence of the elastic energy of sharply
bent 10-nm-size DNA molecules.  \textit{Phys. Rev. E},  \textbf{87}, 022710, 2013.

\bibitem{menon13}
R. Padinhateeri, G.I. Menon,   Stretching and Bending Fluctuations of Short DNA Molecules.  \emph{Biophys. J.},  \textbf{104},  463-471, 2013.

\bibitem{weber09}
G. Weber, J.W. Essex, C. Neylon,   Probing	the	microscopic	flexibility	of	DNA	from
melting	temperatures.  \textit{Nat. Phys.}, \textbf{5}, 769-773,  2009. 















\end{thebibliography}
\end{document}